\documentclass[%
 aip,
 jcp,
 amsmath,amssymb,
preprint,%
]{revtex4-1}

\usepackage{graphicx}
\usepackage{dcolumn}
\usepackage{bm}

\usepackage[utf8]{inputenc}
\usepackage[T1]{fontenc}
\usepackage{mathptmx}
\usepackage{etoolbox}

\makeatletter
\def\@email#1#2{%
 \endgroup
 \patchcmd{\titleblock@produce}
  {\frontmatter@RRAPformat}
  {\frontmatter@RRAPformat{\produce@RRAP{*#1\href{mailto:#2}{#2}}}\frontmatter@RRAPformat}
  {}{}
}%
\makeatother
\begin{document}


\title[]{Is the Electron Hydrated Through Covalent Sharing?}
\author{Y. Sajeev}
\email{sajeevy@barc.gov.in} 
\affiliation{ Bhabha Atomic 
Research Centre, Mumbai-400085, India}

\date{\today}
\begin{abstract}
The hydrated electron ($e_{aq}^-$), a key species in radiation chemistry, is traditionally modeled as an interior electron confined within a solvent cavity and stabilized by electrostatic interactions. However, this picture fails to account for its high binding energy and discrete excited states, as the cavity lacks sufficient dipole strength to support deep electronic confinement. Using \textit{ab initio} methods that capture resonant interactions between the free electron and water, we show that the hydrated electron is stabilized through covalent delocalization. Existing approaches misrepresent this as electrostatic trapping within a cavity---an interpretation rooted in assumptions of a pre-bound electron and the omission of the resonant character of the initial interaction between the free electron and water. Our results reveal that the electron forms transient negative ion molecular states through resonant attachment to neighboring water molecules, where it is initially captured, and delocalizes over them via an intermolecular bonding network formed by the superposition of $a_1$ valence orbitals.  This covalent delocalization yields cavity-like structures without requiring electrostatic trapping and naturally explains observed spectral features, including higher-nodal excited states and enhanced binding energies. In short, cavity formation is initiated by \textit{associative electron attachment (AEA)}---a molecular process driven by a resonant interaction between the free electron and its neighboring water molecules, and wherein the electron becomes covalently shared to them---during the energy dissipation phase of the free electron preceding full solvation.
\end{abstract}

\maketitle

The hydrated electron ($e_{aq}^-$), a free electron stabilized in liquid water, represents the simplest quantum solute~\cite{QSOLUTE} and plays a central role as a reducing agent in radiation chemistry.\cite{HE1, RADCHEM1,RADCHEM2} Speculated over a century ago~\cite{SPECULATION} and experimentally confirmed more than six decades ago,~\cite{ABSORPTION,ABSORPTION1} the hydrated electron continues to defy a complete microscopic description, with experimental findings still lacking a definitive theoretical explanation.~\cite{HE1,DEBATE1,WETE}
The most widely accepted model envisions the electron confined within a quasi-spherical cavity formed by surrounding water molecules, stabilized through weak electrostatic interactions with the partial positive charges of nearby hydrogen atoms.~\cite{CAVITY1,CAVITY2,CAVITY3,DEBATE2,DEBATE3} Within this framework, the ground-state wavefunction of the electron resembles a nodeless, \textit{s}-like orbital, while the photo-excited states resemble \textit{p}-like orbitals, analogous to a free electron confined in a spherical potential well.~\cite{HE1,SP,SP1,SP2}

Modern experimental observations provide partial support for the cavity-based picture of the hydrated electron. Optical absorption spectra reveal bound-bound electronic transitions, suggesting discrete energy levels consistent with spatial confinement.~\cite{SP,SP1,SP2} X-ray absorption spectroscopy (XAS) further supports the existence of a cavity,~\cite{XAFS} with time-resolved XAS (trXAS) capturing its dynamic formation.~\cite{ROBINJACS} However, several critical inconsistencies undermine the completeness of this model---most notably, the high binding energy of $\sim$3.7 eV,~\cite{BE,BE1} which cannot be explained by weak electrostatic interactions alone. Moreover, the proposed cavity structure lacks a net-dipole-moment, casting doubt on its ability to support multiple bound excited states. Most significantly, simulations that reproduce the results of state-of-the-art X-ray absorption spectroscopy experiments reveal substantial contributions from valence orbitals to the excitation transitions~\cite{XAS}---features that cannot be explained within the conventional electrostatic confinement framework of the cavity model. These discrepancies strongly suggest that a purely electrostatic description is insufficient, and that electron stabilization in water likely involves more fundamental molecular-level interactions.

To address these discrepancies, we revisit the fundamental interaction between a free electron and individual water molecules. This interaction is intrinsically quantum mechanical, governed by low-energy scattering processes that can become resonant at specific electron energies. Under such conditions, the electron can be  transiently  captured  by the  electronic field of nearby water molecules, forming short-lived but structurally significant electron-water molecular compound states.~\cite{EK,RADCHEM2,SANCHE1} Although these resonant states are well established in radiation chemistry and low-energy electron scattering,~\cite{RADCHEM1,RADCHEM2,SANCHE1} their role in hydrated electron formation has not been adequately explored. We hypothesize that such resonant electron-water interactions, accessible during the energy dissipation phase of free-electrons in water  preceding solvation, serves as  essential precursors to cavity formation. Since larger neutral water clusters exhibit a positive vertical electron affinity,~\cite{VITJA} with electron attachment directly yielding the ground state of the solvated electron, these larger neutral clusters are unlikely to engage in resonant interactions with a free electron to produce that same ground state. As a result, only small molecular aggregates are capable of sustaining  such resonant interactions, leading to the formation of the required negative-ion ground state of the microhydrated electron. To be more precise, we propose that free electrons undergo resonant interactions with minimally sized, locally coordinated water molecules---interactions that play a critical role in triggering structural reorganization and ultimately leading to the formation of cavity-like configurations that stabilize the hydrated electron.

To uncover the stabilization mechanism originating from the resonant interaction using  first principles, we  adopt a bottom-up  quantum chemical approach---beginning with isolated water molecules and progressively  increasing the  environmental complexity to sustain such resonances, leading to the formation of transient electron-molecular compound states that may act as precursors to the cavity-stabilized hydrated electron.  Specifically, our investigation centers on small water clusters containing up to four molecules, which represent the minimal configurations known to thermodynamically stabilize an excess electron.~\cite{VITJA} Using \textit{ab initio} quantum chemical methods, we model these electron-water compound states as true bound anions rather than continuum resonances. To achieve this, we employ the Coupled-Cluster method with single and double excitations, in conjunction with the \textit{spatially compact} 6-31G Slater-type Gaussian basis set. This methodological framework enables a reliable qualitative characterization of the key structural and electronic features of transient electron-molecule compound states, in which the \textit{spatially compact basis set} plays a critical role in capturing the resonant covalent interaction---a point to which we will return in our discussion of the results.  All calculations are performed using the GAMESS-US quantum chemical code.~\cite{GAMESS} The results are presented in Figures 1 and 2, and their broader implications for understanding the molecular origin of the cavity-stabilized hydrated electron are discussed below.

In the lowest energy regime, a free electron resonantly interacting with an isolated water molecule is captured into the lowest unoccupied molecular orbital (LUMO) of that molecule--the $a_1$ valence orbital (see Fig. 1)--forming a transient one-particle electron-water compound state with valence character.  This transient negative ion compound state represents the initial step in the formation of a molecular anion complex and serves as the quantum-chemical foundation of our investigation. It is important to clarify, for the purposes of the following discussion, that although the dipole-moment of water facilitates this capture, the resulting state is best described as a dipole-supported electron-molecule compound state with valence character, rather than a true dipole-bound anionic state, which would involve binding to diffuse, non-valence orbitals. Due to its strong $\sigma^*$ character and spatial localization, this state rapidly decays via autodetachment in the absence of additional water molecules.~\cite{SIGMASTAR}

A remarkable transformation occurs when a second water molecule enters the interaction region of a free electron. In this configuration, the electron can be resonantly captured and covalently delocalized between the localized $a_1$ valence orbitals of the two water molecules, resulting in a transient, delocalized negative-ion compound state with valence character. This covalent delocalization enhances both the energetic stability and lifetime of the transient compound, opening new pathways for structural relaxation and the formation of more stable negative ion species.
As additional water molecules participate, this transient state evolves into a thermodynamically stable network of water molecules linked through electron-mediated covalent interactions.  This process, known in molecular physics as associative electron attachment (AEA),~\cite{AEA1,AEA2} is represented by:
\begin{equation} e^-(\epsilon) \hskip 20pt + \hskip 20pt  (H_2O)_1 + (H_2O)_2 + \dots \hskip 20pt \to \hskip 20pt [(H_2O)_1+(H_2O)_2+\dots]^- \end{equation}
The excess electron occupies an intermolecular bonding orbital given by:
\begin{equation} \phi_{AEA} = c_1\phi_{w1}(a_1) + c_2\phi_{w2}(a_1) + \dots \end{equation}
where $\phi_{w}$ is a superposition of the a$_1$ valence orbitals of the participating water molecules. This intermolecular bonding network, arising from the quantum superposition of localized valence orbitals, stabilizes the excess electron through delocalization and introduces a fundamentally distinct quantum-chemical mechanism for confinement and hydration. Rather than being electrostatically confined within a cavity, the electron---via resonant interactions---forms a covalent compound state with surrounding water molecules, stabilized through delocalization over molecular valence orbitals.

Our quantum chemical calculations yielded geometrically optimized, minimum-energy structures of electron-molecule compound states, obtained by allowing water molecules to relax in the ground-state configuration of the \textit{one-particle} negative ion state--i.e., molecular associates formed via the associative electron attachment (AEA) process. Remarkably, these covalent assemblies exhibit cavity formation upon the successive addition of water molecules, as shown in Figure 1. Although the resulting cavity lacks a net-dipole-moment---thereby ruling out conventional electrostatic stabilization---the electron remains confined, stabilized instead through covalent orbital delocalization. As discussed above, this confinement originates from the resonant occupation of an intermolecular orbital formed by the covalent superposition of $a_1$ orbitals from neighboring water molecules. Rather than recognizing this covalently delocalized state as a result of resonant interaction, all of the earlier interpretations overlooked the central, nodeless region of the bonding orbital and mistakenly identified it as a simple \textit{s}-like orbital of an internally trapped free electron,~\cite{INTERIOR} with excited states misassigned as  \textit{p}-like states.

In this context, it is also worth briefly discussing why our simplified CCSD/6-31G approach succeeds where conventional methods remain unpredictable, as well as addressing the inherent issues associated with larger basis sets and how these can be avoided. Although conventional \textit{ab initio} and QM/MM methods can reproduce the final cavity structure of the hydrated electron, they are intrinsically incapable of capturing the underlying mechanism responsible for its formation, and have therefore misconceptualized the electron as an internally trapped particle of the cavity. Electron hydration in radiation chemistry begins with a scattering event, in which a free electron interacts with nearby molecules and becomes bound. This fundamental process--governed by free-electron-molecule interactions---lies outside the scope of conventional \textit{ab initio} quantum chemical and QM/MM methods, which are confined to treating bounded electrons  within the Hermitian framework on which they are built. Moreover, when modeling the free-electron continuum, they rely on large, diffuse basis sets within a Hermitian formalism that inherently discretize the continuum and fail to capture the dominant resonant, covalent interactions. As a result, they are intrinsically blind to the covalent, resonance-driven interactions between the electron and surrounding water molecules. Consequently, while they  produce the final cavity-like structure, which is a bound state,  they cannot relate it to its true origin: the covalent framework formed through associative electron attachment. Therefore, in this work, we validate the covalent origin of electron hydration  directly with experimental observations and deliberately avoid relying on large-scale \textit{ab initio} or QM/MM data for validation.

In contrast, the \textit{spatially compact basis set} employed in our approach captures the resonant electron-molecule interaction without being lost in the large, discretized continuum introduced by diffuse basis sets, and thereby reveals the stabilization mechanism arising from covalent sharing with neighboring water molecules. This localized, quantum-mechanical treatment--though based on the simplest \textit{ab initio} framework that captures the resonant interaction between free-electron and molecules--offers a physically grounded alternative to the bulk descriptions of the hydrated electron generated by conventional \textit{ab initio} and QM/MM approaches.
Our use of the CCSD/6-31G compact basis set approach may be misunderstood or underestimated. However, it is sufficient to qualitatively capture the covalent nature of the transient negative ion formed between the free electron and water molecules, using a uniquely defined wavefunction within a Hermitian framework, as implemented in widely available quantum chemical packages.

For quantitative accuracy, one could employ a larger basis set. However, obtaining a similarly well-defined wavefunction for transient negative ion states---one that reveals the covalent character--within the Hermitian quantum chemical framework typically requires stabilization techniques~(see Ref. \citenum{STABI}, and more references therein). Alternatively, with large basis sets and still within the Hermitian framework, the covalent features of the transient negative ion can also be described using continuum-remover potentials~\cite{CRCAP,RVCR}. For the most rigorous treatment, a large basis set combined with a non-Hermitian quantum chemical framework allows for a unique and accurate characterization of the transient negative ion and its covalent bonding features~\cite{NIMRODBOOK,LENZ1,DIETER,RFCAP,RFH2,FOCK,FPO,AK}.
Without such methodological choices on the quantum mechanical  side of QM/MM approaches, the resonant interaction of the free electron may appear merely coincidental. Consequently, only those QM/MM strategies capable of sustaining a cavity-forming model may retrospectively reinterpret the results to provide a qualitative estimate of the relative contributions of electrostatic versus covalent effects in the hydration of the free electron.

\begin{figure}[!h]
\label{stab1}
\centering
\includegraphics[scale=0.8]{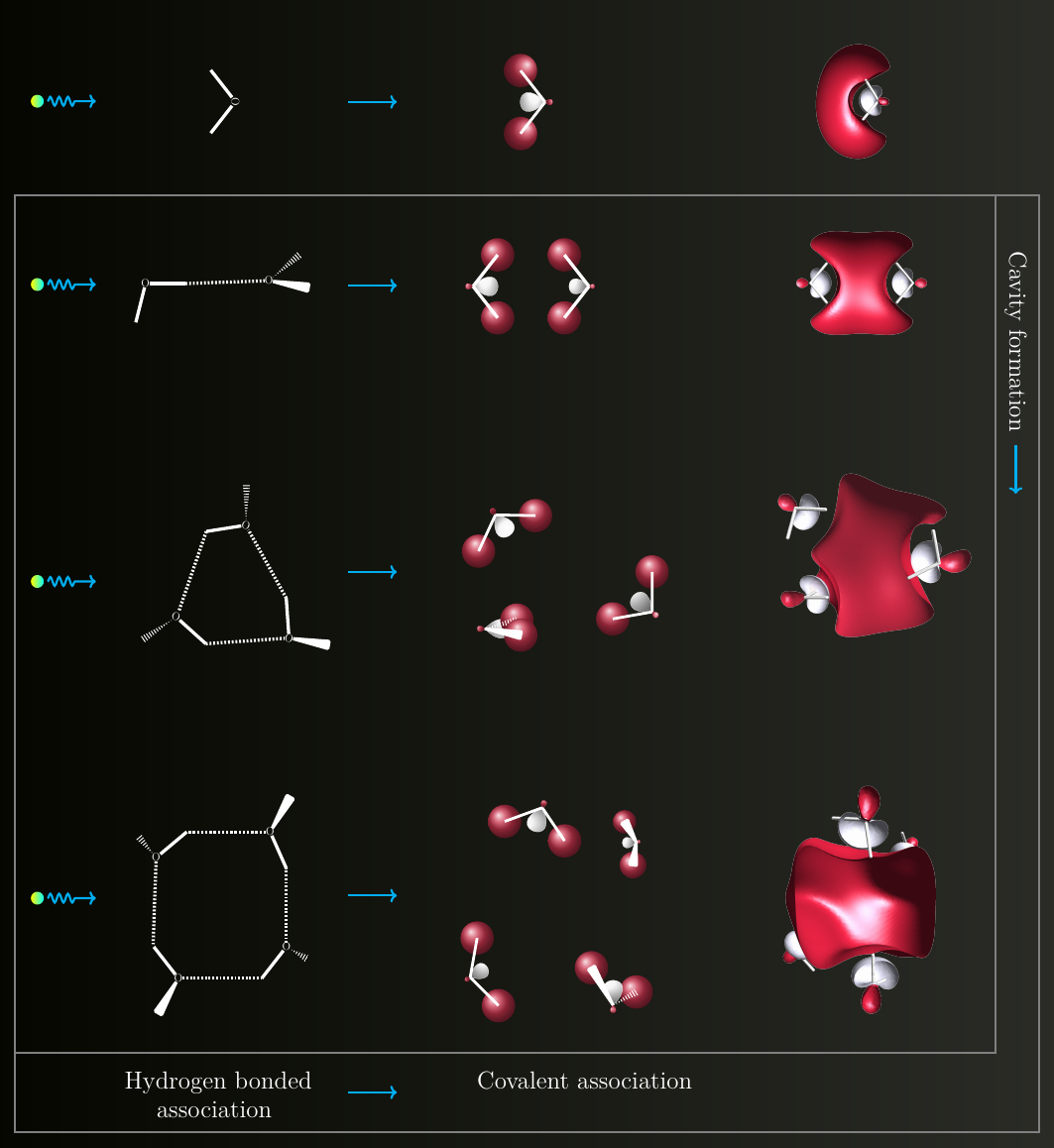}
\vskip 5pt 
\caption{ Resonant capture of a free electron by a hydrogen-bonded water cluster initiates covalent association among the water molecules, subsequently leading to cavity formation, as illustrated. The localized $a_1$ orbitals of the individual water molecules are shown. The orbital responsible for covalent binding of the electron---corresponding to the lowest-energy superposition of these $a_1$ valence orbitals---is also depicted. For reference, the upper panel shows the resonant capture of an electron into the $a_1$ valence orbital of an isolated water molecule. }
\vskip -10pt 
\end{figure}

The covalent nature of water association through resonant interaction is further demonstrated in our calculations of a hypothetical transient di-negative ion state,  where a second free electron is resonantly captured to the intermolecular bonding orbitals of the minimum-energy negative-ion structures shown in Figure 1. As illustrated in Figure 2, this second electron enhances the intermolecular bond order, resulting in a more compact and highly symmetric cavity---a direct manifestation of covalent association. While this result is consistent with the traditional cavity model, it highlights that covalent interaction is the fundamental driving force behind cavity formation. Further analysis of microhydrated electrons and dielectrons reveals that higher-energy discrete negative ion states also arise naturally from intermolecular orbitals with increasing nodal structure, formed through the superposition of individual water molecule $a_1$ orbitals, as shown in Figure 2.

As further support for the covalent nature of electron hydration proposed in this work, the highly accurate energy and structural data reported by the groups of Prof. Cederbaum and Prof. Jordan~\cite{VITJA} offer valuable benchmarks for electron-molecule compound states, particularly those involving interior electrons confined within nonpolar water cavities. Although their focus was on establishing benchmark-level energetics, their \textit{ab initio} calculations inherently capture covalent interactions, offering a meaningful point of comparison for our investigation into the covalent nature of electron binding---a central aspect of understanding electron hydration. They report a binding energy of approximately 800 meV for an interior excess electron stabilized within a cavity formed by as few as six water molecules, increasing to nearly $\sim$1 eV for eight molecules. These findings underscore the critical role of molecular-scale covalent stabilization in explaining the high experimental binding energy of ~3.7 eV observed for the solvated electron in bulk water. While the cavity structure originates from localized covalent interactions at the molecular level, it is ultimately stabilized as the solvated electron through the collective influence of the outer bulk water molecules. This final stabilization arises from long-range electrostatic forces, many-body interactions, and bulk solvent effects, which together ensure the strong binding of the excess electron. It is important, however, not to conflate this bulk-mediated stabilization with the origin of the cavity itself: the spectroscopic features, strong initial binding, and distinct molecular reactivity all trace back to covalent and resonant interactions between the free electron and individual water molecules.

\begin{figure}
\label{stab2}
\centering
\includegraphics[scale=1]{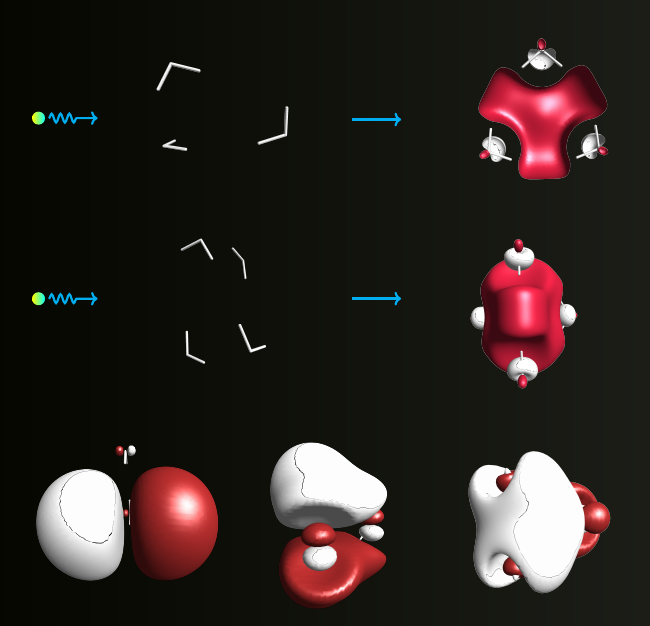}
\vskip 5pt 
\caption{ The strong covalent association and the formation of increasingly compact and symmetric cavity structures in a hypothetical transient di-negative ion state---arising from the resonant capture of a second free electron into the transient negative ion states of trimeric and tetrameric water cluster---is illustrated.   The corresponding electron densities---arising from higher-nodal superpositions of the $a_1$ valence orbitals of the constituent water molecules---are shown below, illustrating the emergence of three excited, $p$-like states within the tetrameric water cavity.}
\vskip -10pt 
\end{figure}

Before proceeding to the conclusion, it is worth noting that recent time-dependent DFT simulations of the X-ray absorption spectrum of the hydrated electron, carried out within a QM/MM framework, indicate the involvement of the water $a_1$ orbital.~\cite{XAS} However, like many other quantum chemical studies of the hydrated electron, this work overlooks the apparent associative electron attachment (AEA) pathway, as well as the roles of molecular orbital superposition and covalent delocalization. Notably, a pre-hydrated dipole-bound electron with very low binding energy---undergoing a transition into a localized, cavity-like state that disrupts the hydrogen-bond network, and reportedly reverting back to the dipole-bound configuration---has also been reported recently.~\cite{WETE} However, this work did not examine the valence character of this localized state, nor did they extend their interpretation of the  pre-heydrated electron toward the resonant regime of the free electron---once again reflecting the limitations of Hermitian frameworks and conventional quantum chemical approaches in capturing resonant electron-molecule interactions.

In summary, our work proposes a covalent, resonant mechanism for hydrated electron formation, challenging the traditional view of electrostatic trapping within a solvent cavity. As the free electron dissipates its energy in water, it engages in resonant interactions with nearby molecules, enabling the occupation of bonding orbitals formed through the overlap of water $a_1$ valence orbitals---interactions that naturally give rise to cavity-like structures through covalent delocalization. That is,  cavity formation is triggered by the resonant, covalently mediated \textit{associative electron attachment (AEA)} of the free electron to its neighboring water molecules during the energy dissipation phase prior to full solvation. Dielectron calculations further support this framework, revealing enhanced bond order and geometric symmetry consistent with covalent association. Most significantly, covalent delocalization directly addresses the valence orbital contributions from water molecules observed in electronic excitation transitions, as revealed by modern X-ray absorption studies. Thus, covalent delocalization not only reconciles the cavity model with quantum chemical insights but also  unifies the structural motifs, energetic stability, and spectroscopic fingerprints of the hydrated electron.

\bibliography{HYDRATED} 
\end{document}